\documentclass[12pt]{iopart}
\usepackage{iopams}  
\usepackage{graphicx}
\def\be{\begin{equation}}
\def\ee{\end{equation}}

\begin{document}

\title{ 
Meson correlators above deconfinement
}

\author{Saumen Datta}
\address{
Fakult\"at f\"ur Physik, Universit\"at Bielefeld, 
D33615 Bielefeld, Germany.}
\author{Frithjof Karsch}
\address{
Fakult\"at f\"ur Physik, Universit\"at Bielefeld, 
D33615 Bielefeld, Germany}
\author{P\'eter Petreczky 
\footnote{speaker at the conference}}
\address{ 
Nuclear Theory Group, Department of Physics, Brookhaven
National Laboratory, Upton, New York, 11973
}
\author{Ines Wetzorke}
\address{NIC/DESY Zeuthen, Platanenallee 6, D-15738 Zeuthen, Germany.}

\begin{abstract}
We review recent progress in studying spectral functions
for mesonic observables at finite temperatures, by analysis 
of imaginary time correlators directly calculated on isotropic
lattices. Special attention is paid to the lattice artifacts
present in such calculations.
\end{abstract}

\pacs{11.15.Ha, 11.10.Wx, 12.38.Mh, 25.75.Nq}

\submitto{\JPG}


\section{Introduction}
Spectral functions of mesonic operators play an important role
in finite temperature QCD. Many experimental results
in high energy heavy ion collisions ( e.g. low invariant mass
dilepton enhancement, anomalous $J/\psi$ suppression etc.)
can be understood in terms of medium modifications of meson
spectral functions \cite{rapp}.

For a long time it has been believed 
that lattice QCD is suitable only for calculation
of static quantities at finite temperatures, 
such as the transition temperature,
equation of state, screening lengths etc. However, it was shown by
Asakawa, Hatsuda and Nakahara that using the {\em Maximum Entropy Method}
one can in principle reconstruct also meson spectral functions.
The method was successfully applied at zero temperature 
\cite{nakahara99,yamazaki02,blum04}.
and also at finite temperature
\cite{karsch02dil,karsch03qm,asakawa03qm,petreczky03sqm,umeda02,asakawa03hq,datta04}.
Though systematic uncertainties in 
the spectral function calculated on lattice are not yet completely 
understood, it was shown in Ref. \cite{karsch02dil} that precise determination 
of the imaginary time correlator can alone provide constraints 
on the spectral function at finite temperature. 

In this contribution we are going to review recent results on
meson correlators and spectral functions in finite temperature QCD.

\section{Meson spectral functions}
The spectral function for an operator is directly related
to the Fourier transforms of real time correlators of the
operator. Defining the real time propagators
\begin{equation}
D^{<}(t)=\langle \hat O(0) \hat O(t) \rangle,~
D^{>}(t)=\langle \hat O(t) \hat O(0) \rangle,
\label{eq.realtime}
\end{equation}
for a suitable operator $O$, the spectral function
$\sigma(\omega)$ is defined as 
\begin{equation}
\sigma(\omega)=\frac{D^{>}(\omega)-D^{<}(\omega)}{ 2 \pi},
\end{equation}
where $D^{>,<}(\omega)$ are the Fourier transforms of the
correlators in Eq. (\ref{eq.realtime}). The spectral
function encapsulates the properties of the channel
corresponding to operator $O$: a stable mesonic state will
show up as a $\delta$ function and an unstable state will
be present as a broad (Breit-Wigner) peak. Also, other
quantities of experimental interest can be directly
extracted from spectral functions:
for example, the rate of emission of thermal dileptons can be 
expressed in terms of the spectral function of
the vector current $J_{\mu}=\bar q \gamma_{\mu} q$ \cite{braaten90}
\begin{equation}
{dW \over dp_0 d^3p}  = {5 \alpha_{em}^2 \over 27 \pi^2} 
{1 \over p_0^2 (e^{p_0/T}-1)} \sigma_V(p_0, \vec{p}).
\end{equation}

In lattice studies of finite temperature systems in
equilibrium, on the other hand, one calculates the imaginary time or
Matsubara correlator 
\begin{equation}
G(\tau) = \langle \hat O(\tau) \hat O(0) \rangle,
\label{eq.matsubara} \end{equation}
where the Euclidean time $\tau \in (0, \beta=1/T)$. 
The imaginary time correlators is an analytic continuation of the
real time correlator $G(\tau) = D^{>}(-i \tau)$.
This and the periodicity in $\tau$ lead to an integral
representation for $G(\tau)$ in terms of the spectral function, 
\begin{eqnarray}
G(\tau, T) &=& \int_0^{\infty} d \omega
\sigma(\omega,T) K(\tau,\omega,T) \label{eq.spect} \\
K(\tau,\omega) &=& \frac{\cosh(\omega(\tau-1/2
T))}{\sinh(\omega/2 T)}.
\label{eq.kernel}
\end{eqnarray}
Here we consider correlators of point meson operators
$\bar q \Gamma q$, with 
$\Gamma=1,~\gamma_5,~\gamma_{\mu}, \gamma_{\mu}\gamma_5$ for 
scalar, pseudo-scalar, vector and axial-vector channels.
We will also restrict the discussion to the case of zero 
spatial momentum.

Equation (\ref{eq.kernel}) is valid in the continuum. It is 
not obvious that it holds also for the correlators calculated on the lattice. 
However, it was shown in that an integral representation like
in Eq. (\ref{eq.kernel}) can be also defined for
the lattice correlator in the limit of the free field 
theory, with the only modification being that the upper 
integration limit becomes
finite $\omega_{max} \simeq 4 a^{-1}$ \cite{karsch03fspf}. 
The effect of the discretization is contained solely 
in the spectral function\cite{karsch03fspf}.

In extracting $\sigma(\omega)$ from the lattice
measurements of $G(\tau)$ using Eq. (\ref{eq.kernel}), the obvious
problem is that we have to reconstruct several hundred degrees of freedom
(needed for a reasonable discretization of the integral 
in Eq. (\ref{eq.kernel})
with ${\cal O}(10)$ data point on $G(\tau)$. This ill-posed problem can be 
solved using the Maximum Entropy Method (MEM) \cite{nakahara99}. 
In the Maximum Entropy Method on searches for spectral function which
maximizes the conditional probability $\exp(-\chi^2/2+\alpha S)$,
where $\chi^2/2$ is the standard likelihood function for the data and
\begin{equation}
S=\int_0^{\infty} d \omega \biggl[
\sigma(\omega)-m(\omega)-m(\omega)\ln\frac{\sigma(\omega)}{m(\omega)} \biggr]
\end{equation}
is the Shannon-Jaynes entropy which parametrizes prior knowledges such
as positivity of the spectral functions and informations
about the asymptotic form of the spectral function. 
Finally $\alpha$ is a real and positive parameter, which
is integrated out to obtain the most probable spectral function.
The Shannon-Jaynes entropy depends on one real function
$m(\omega)$ called the default model.

\section{Numerical results for the correlators and spectral functions}

Here we will mostly discuss finite temperature
spectral functions for heavy quarkonia. The behavior of
heavy quarkonia at finite temperature has been of great
interest ever since the suggestion of Matsui and Satz
that the disappearance of $J/\psi$ can 
act as a signal of onset of deconfinement, since the 
screening in a quark-gluon plasma will lead to the
dissolution of charmonium bound states \cite{MS86}.
The problem of the $J/\psi$ suppression was addressed using potential
models \cite{MS86,karsch88,digal01}, and resulted in
estimates that the $J/\psi$ will dissolve around $1.1T_c$ 
\cite{karsch88,digal01}. More recent calculations, using
lattice results for internal energy of a static $\bar{q} q$
pair \cite{okacz04} as the potential, 
lead to a much higher dissociation temperature $> 1.5 T_c$
for the $J/\psi$ \cite{shuryak04,wong04}.

Direct lattice investigations of heavy quarkonia at finite
temperatures, by studying the Matsubara correlators and
extracting the spectral funtion from them using MEM, were
performed only recently.
Calculations have been done using anisotropic 
\cite{umeda02,asakawa03hq} as well
as fine isotropic \cite{datta04} lattices in quenched
approximation (i.e., for a purely gluonic plasma).
A check of the cutoff dependence of the results by using
several lattice spacings, which turns out to be quite important, 
has been done so far only in Ref. \cite{datta04}.
The extracted spectral functions at zero temperature 
reproduced the properties of the ground state mesons, but  
revealed structures (lattice artifacts) at higher energies whose 
positions scale like $a^{-1}$ \cite{yamazaki02,blum04,datta04}. 
Due to these structures, the high energy part of the
spectral functions in the interacting theory look very
different from that in the free theory.

\begin{figure*}
\includegraphics[width=2.5in]{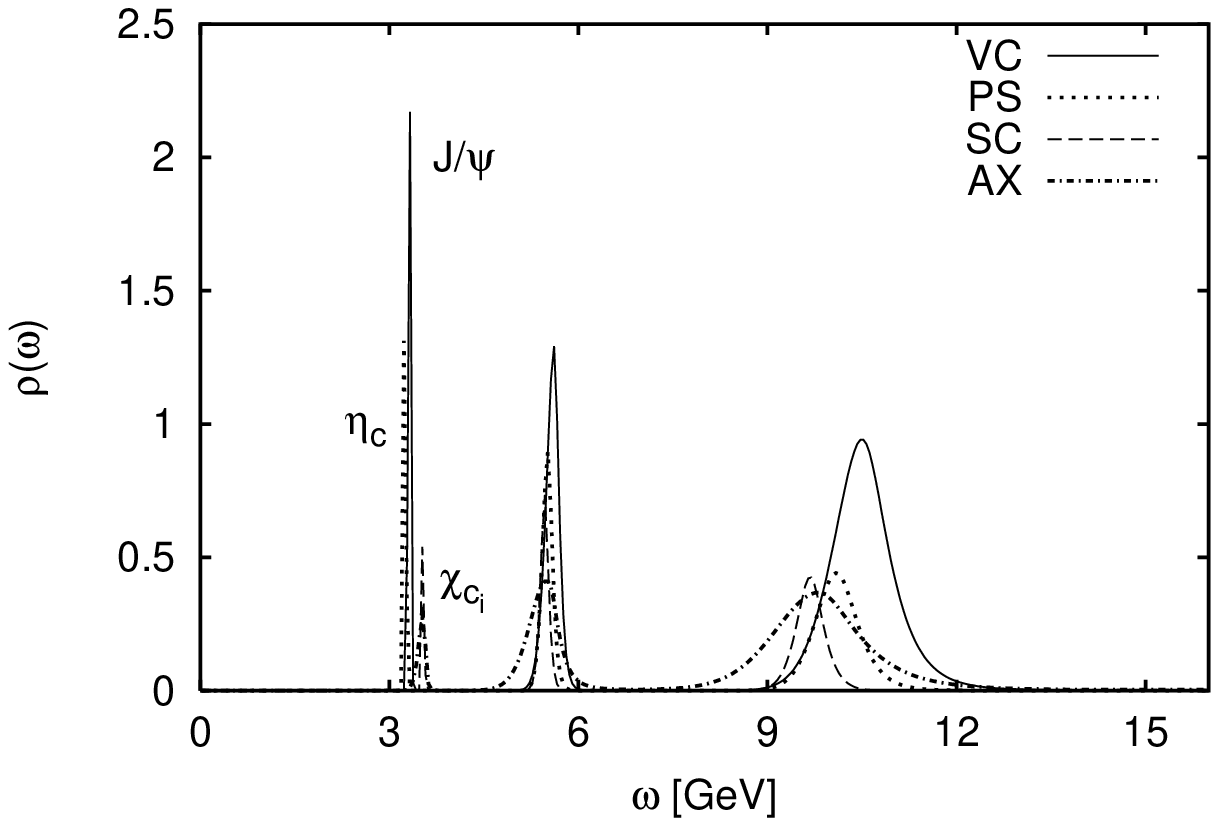}
\includegraphics[width=2.5in]{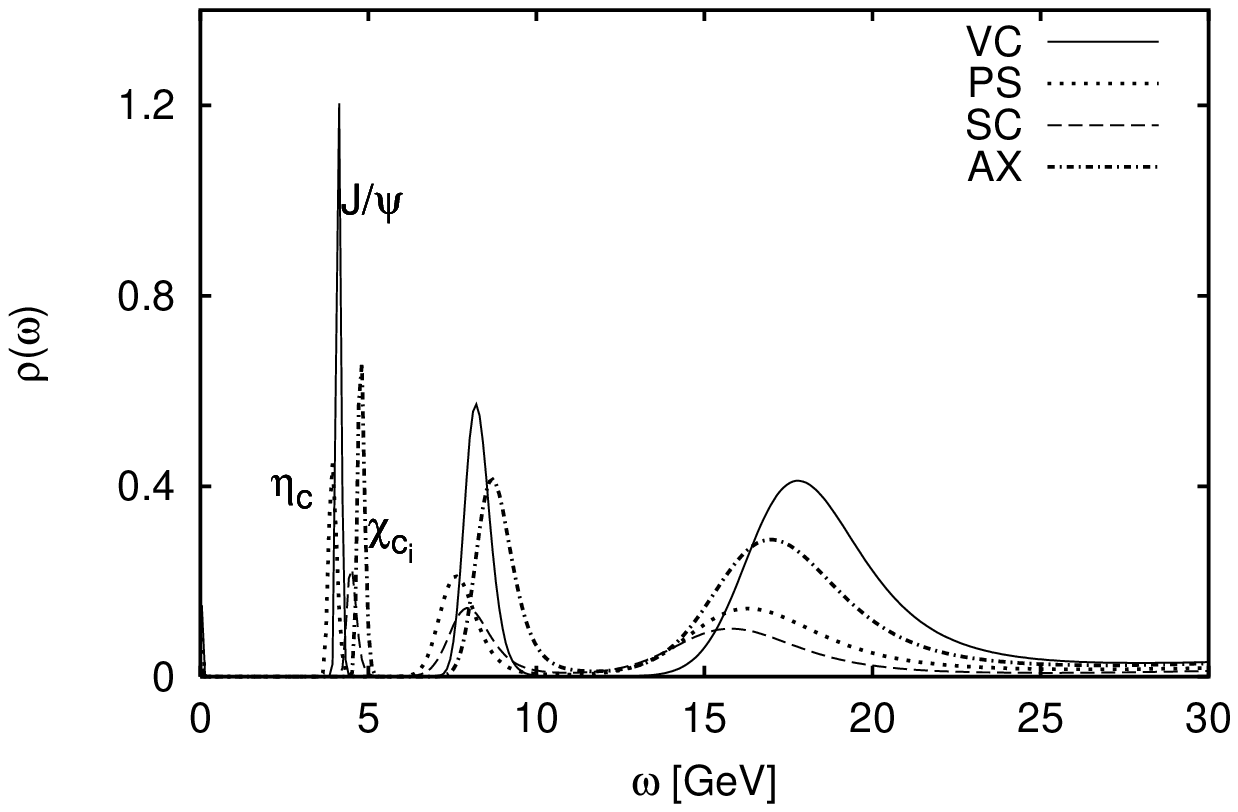}
\vspace*{-0.6cm}
\caption{The charmonia spectral functions below deconfinement
for two different lattice spacings $a^{-1}=4.86GeV$ (left)
and $a^{-1}=9.72GeV$ (right) \cite{datta04}.}
\label{comp_a_spf}
\vspace*{-0.5cm}
\end{figure*}

We first discuss the spectral functions at low
temperatures where the large extent of the imaginary time
direction allows a reliable extraction of the spectral
function from the Matsubara correlators. We have used the 
continuum free meson spectral
function of the form $m_0 \omega^2$ as a default model. 
It was checked that the result does not depend on the choice
of the default model \cite{datta04}.
In Fig. \ref{comp_a_spf} we show
charmonium spectral functions below deconfinement at two different
lattice spacings, corresponding to temperatures of 
$0.75T_c$ and $0.9T_c$ \cite{datta04}.
The ground state peaks in the pseudo-scalar and vector channels
correspond to $\eta_c$ and $J/\psi$ states, while those in the scalar
and axial-vector channels correspond to $\chi_{c0}$ and
$\chi_{c1}$ states, respectively. While the ground state 
peaks reproduce the expected masses the positions
of the other peaks scales roughly as $1/a$, i.e. these peaks 
are lattice artifacts. 

Now let us discuss what happens at finite temperature.  In this case
the reconstruction of the spectral function becomes more difficult
as not only the number of available time-slices is reduced but also
the extent of the imaginary time direction ($=1/T$) may be quite small
(see discussion in Ref. \cite{umeda02} on this point).
While the former limitation can be relaxed using larger
anisotropies, the latter will always remain. Therefore it is useful
to inspect the temperature dependence of the correlators first.
The temperature dependence of the correlators is due both to
the explicit temperature dependence of the corresponding spectral functions 
and to the temperature dependence of the integration kernel
in Eq. (\ref{eq.kernel}). To factor out the trivial temperature
dependence due to the kernel,we
construct the model correlators
\begin{equation}
G_{rec}(\tau,T)=\int_0^{\infty} d \omega \sigma(\omega,T^{*})K(\tau,\omega,T),
\end{equation}
\begin{figure*}
\includegraphics[width=2.5in]{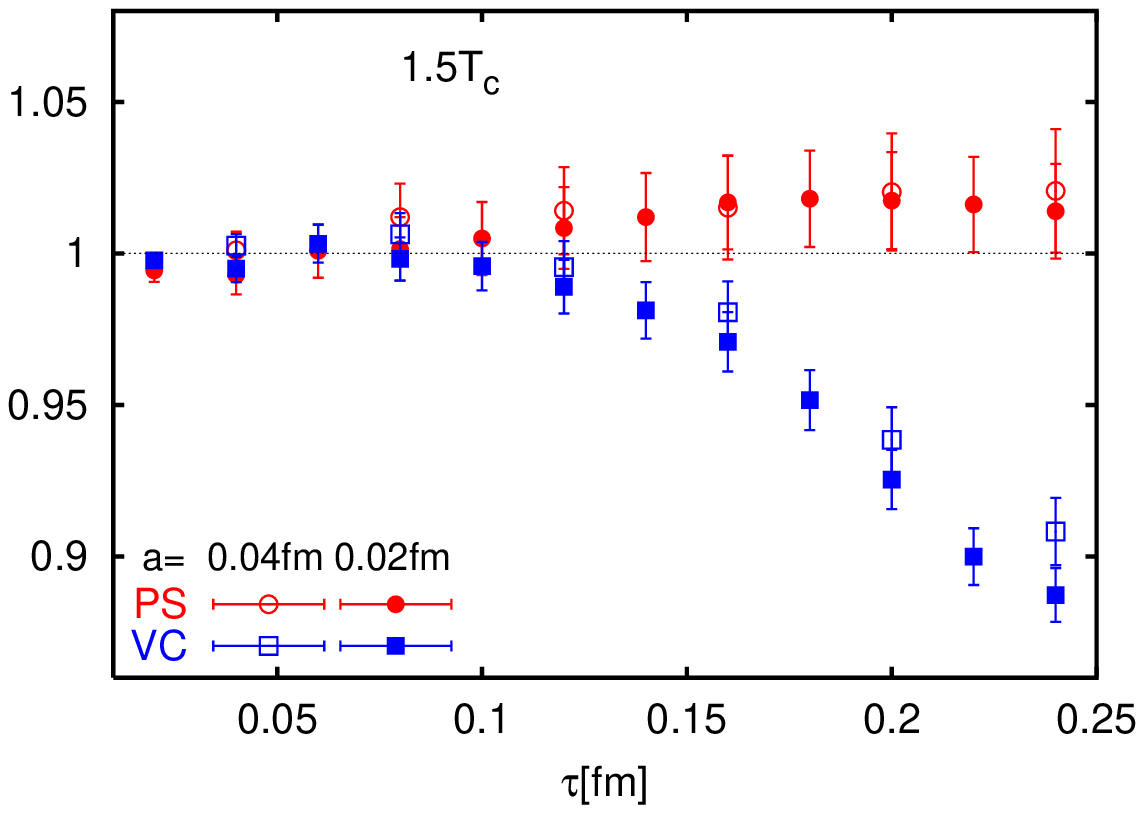}
\includegraphics[width=2.5in]{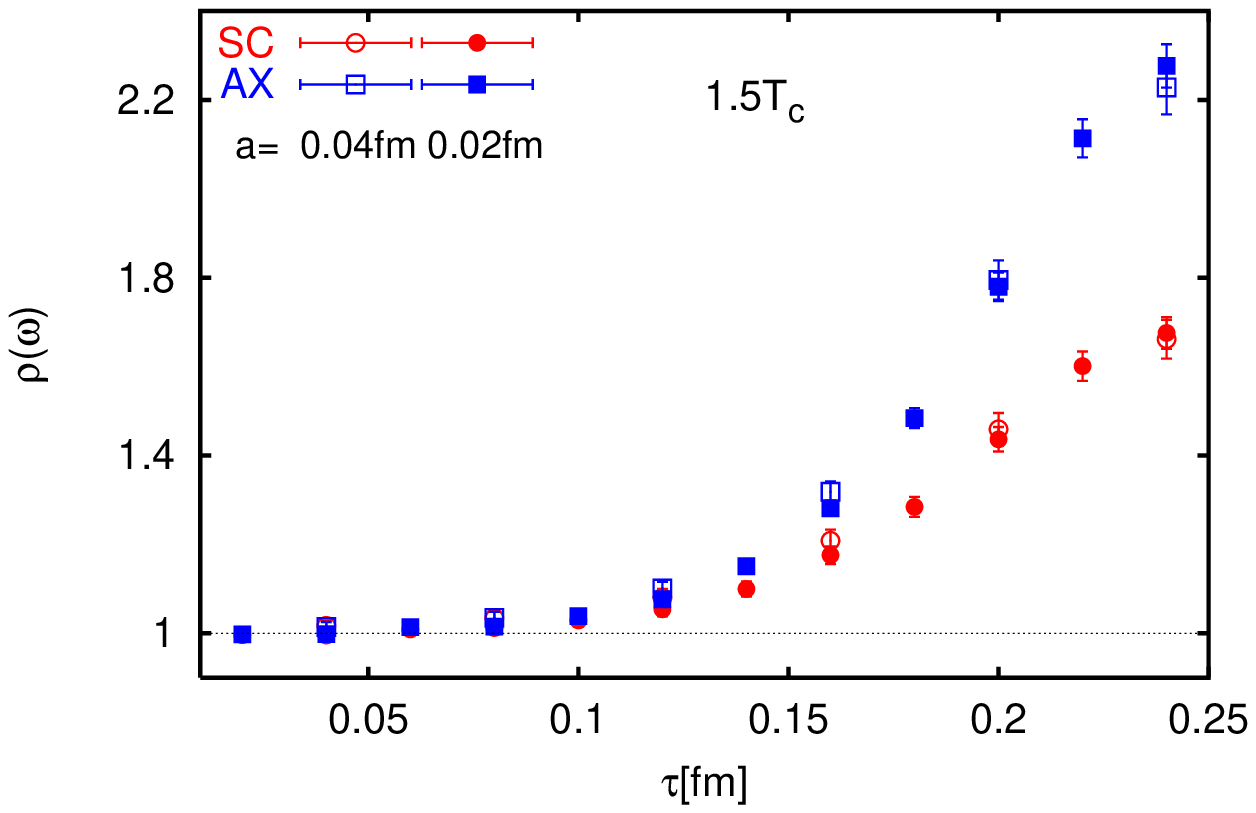}
\vspace*{-0.6cm}
\caption{Ratios $G/G_{recon}$ for the pseudo-scalar and the 
vector correlators (left)
and for the scalar and the axial-vector correlators (right).
}
\label{ratio}
\vspace*{-0.4cm}
\end{figure*}
with $T^{*}$ being some temperature below deconfinement. If the
spectral function has no significant temperature dependence
also above deconfinement, we would expect $G/G_{rec} \simeq
1$ also at temperatures above $T_c$.
In Fig. \ref{ratio} we show $G/G_{rec}$ for different channels at $1.5T_c$.
As one can see
this ratio is not very far from unity for
pseudo-scalar and vector channels, while for the scalar and axial-vector
channel large changes are visible.
In fact, for the  scalar and axial-vector channels large
changes are seen already at $1.1T_c$ \cite{datta04}.
This suggests that the 1S state charmonia ($\eta_c$ and
$J/\psi$) may survive deconfinement with little 
modifications of their properties, while 1P state charmonia 
($\chi_{c_0}$ and $\chi_{c_1}$) are dissolved or strongly
modified by the medium right after deconfinement. 
Notice that Fig. \ref{ratio} shows the ratio $G/G_{rec}$
for two different lattice spacings, with very little
lattice spacing dependence observed in the results.

\begin{figure*}
\includegraphics[width=2.2in]{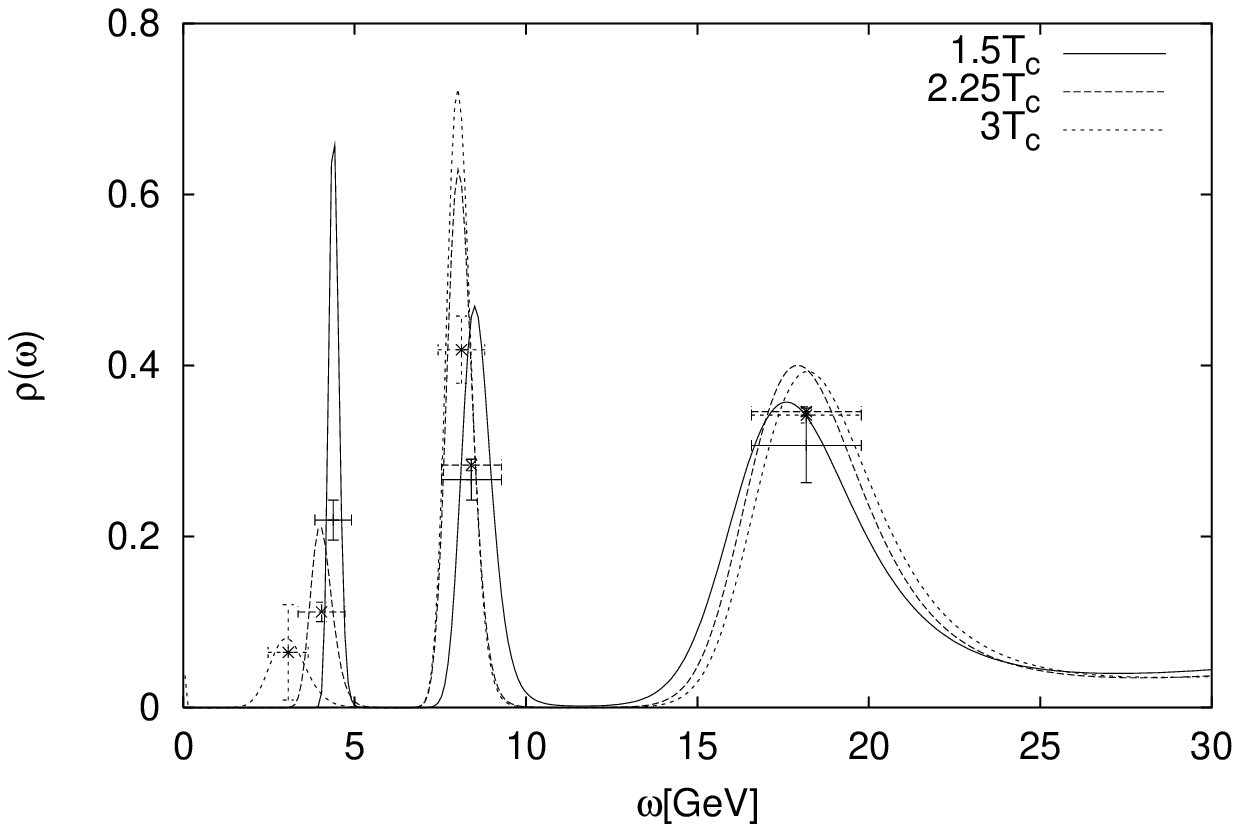}
\includegraphics[width=2.3in]{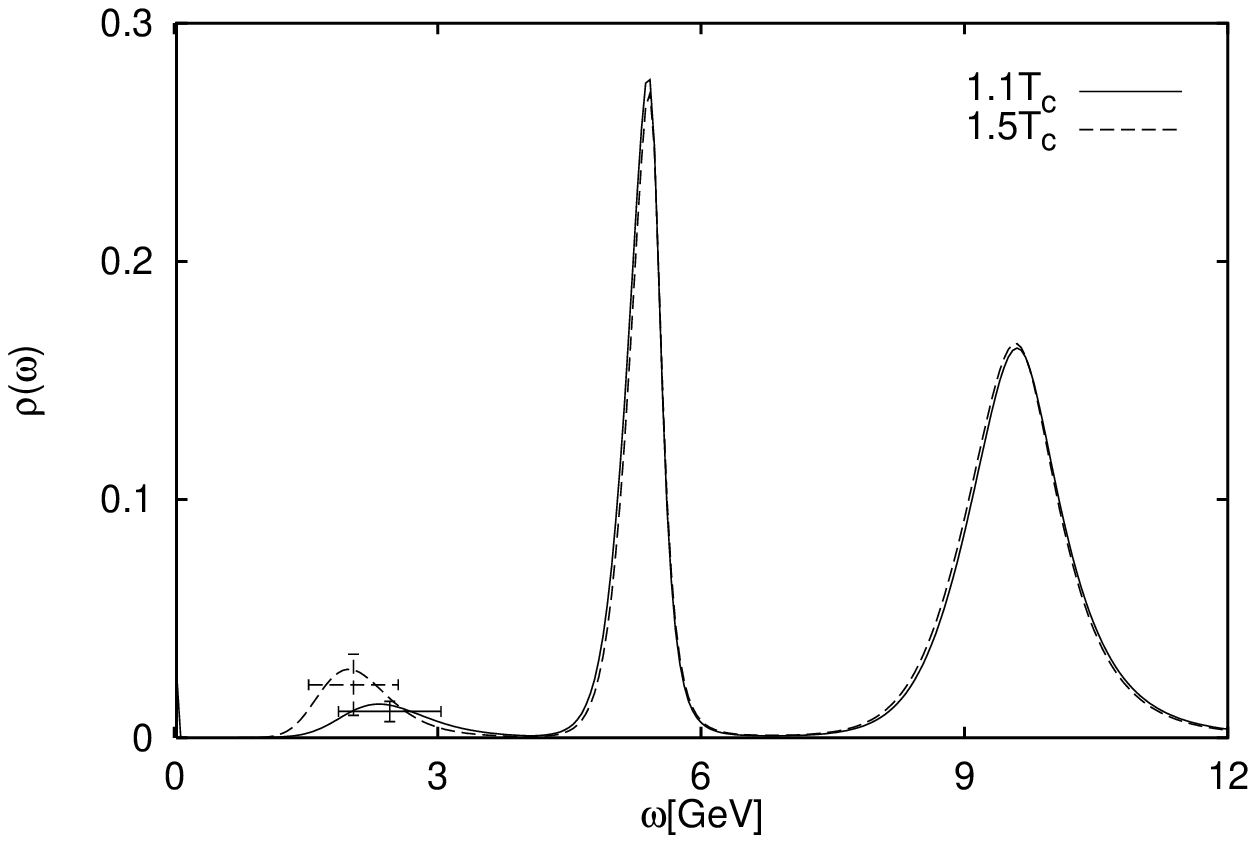}
\vspace*{-0.5cm}
\caption{Charmonia spectral functions for the vector (left) and 
scalar (right)
channels above $T_c$.
\cite{datta04}.}
\label{charm_spf_FT}
\vspace*{-0.6cm}
\end{figure*}

To study the problem in more detail one
needs to extract the spectral functions from the measured correlators.
Because of the limited extent of the time direction this
becomes quite difficuilt at high temperatures, as the result
gets sensitive to the default model. We have used the
high energy part of spectral function below $T_c$
as part of the default model. This is justified as the high
energy part of the extracted spectral function is dominated
by lattice artifacts, and also, the effects of moderate
temperatures should not be significant at very high energies.
In Fig. \ref{charm_spf_FT} we show the spectral functions 
extracted for the vector and scalar channels, at
temperatures of 1.5 -- 3 $T_c$. 
The figure confirms the expectation based on the analysis of
the correlator; the $J/\psi$ is seen to exist up to temperatures
$2.25T_c$ , while the $\chi_{c0}$ state is dissolved
already at $1.1T_c$.
The situation is similar for the pseudo-scalar and axial vector
spectral functions. The results for the survival of the 
$1S$ states till quite high temperatures 
are confirmed by calculations done on anisotropic 
lattices \cite{umeda02,asakawa03hq}, though
there is some controversy about the temperature where $J/\psi$ dissolves
and the way it disappears. In Ref. \cite{asakawa03hq} it was found that the $1S$ charmonia
abruptly disappear at $1.7T_c$, while the spectral functions shown above
suggest a gradual dissolution with resonance structure surviving 
till $2.25T_c$.

\section{Conclusions}
We have discussed lattice calculations of the charmonium spectral functions
at finite temperature. Despite the uncertainties present in the current
analysis we can conclude that $J/\psi$ and $\eta_c$ survive
in the gluonic plasma up to quite high temperatures,
possibly to temperatures as high as $2.25T_c$. 
Moreover, the properties of the 1S charmonia seems to be only
very mildly affected by the deconfinement transition.
The 1P states charmonia on the other hand are dissolved
right after deconfinement. This conclusion is supported both by the
explicit calculations of the spectral functions using MEM
and by the analysis of the correlators which is much more robust. The spectral
functions calculated on the lattice show strong lattice artifacts
at high energies. However the properties of the first peak, which corresponds
to the physical state, are not strongly affected by
these artifacts. This conclusion is also supported by direct
analysis of meson correlators at two different lattice spacings.

Finally we note that meson spectral functions have also been 
analyzed for the case of light quarks 
\cite{karsch02dil,karsch03qm,petreczky03sqm}. Contrary to
the case for the heavy quarks, the spectral functions for
light mesonic observables show quite
strong temperature dependence. However, also here one
sees resonance like structures at low energies and no sign of free
quark propagation. The lattice artifacts at high energy are present
also in this case but were not studied in detail yet. This
would be absolutely necessary for drawing physical
conclusions with confidence.

\ack
This work has been partly supported by 
the U.S. Department of Energy
under contract number DE-AC02-98CH10886. P.P is Goldhaber and RIKEN
Fellow.

\end{document}